\documentclass{aastex631}
\usepackage{graphicx}

\newcommand\mps{Max-Planck-Institut f\"ur
Sonnensystemforschung, Justus-von-Liebig-Weg 3, 37077 G\"ottingen, Germany}

\begin{document}

%\title{Stellar magnetic activity mimics morning and evening asymmetry of planetary terminator}

\title{The effect of stellar magnetic activity on measurements of morning and evening asymmetry of planetary terminator}
%\title{Morning or evening on planet or stellar magnetic activity. What causes the asymmetry of transit light curve?}

%Stellar magnetic activity as an alternative cause of the asymmetry of transit light curves.

\author[0000-0002-6087-3271]{Kostogryz, N.}
\affiliation{\mps}
\correspondingauthor{Kostogryz, N.} 
\email{kostogryz@mps.mpg.de}

\author[0000-0002-8842-5403]{Shapiro, A.I.}
\affiliation{University of Graz, Austria}
\affiliation{\mps}

\author[0000-0001-9355-3752]{Carone, L.}
\affiliation{Space Research Institute, Austrian Academy of Sciences, Schmiedlstrasse 6, A-8042 Graz, Austria}

\author[0000-0001-7696-8665]{Gizon, L.}
\affiliation{\mps}
\affiliation{Institut f\"ur Astrophysik und Geophysik, Georg-August-Universit\"at G\"ottingen, Friedrich-Hund-Platz 1, 37077 G\"ottingen, Germany}

\author[0000-0002-8275-1371]{Helling, Ch.}
\affiliation{Space Research Institute, Austrian Academy of Sciences,  Graz, Austria}
\affiliation{Graz University of Technology,  Graz, Austria}

\author[0000-0003-1285-3433]{Kiefer, S.}
\affiliation{Department of Astronomy, University of Texas at Austin, 2515 Speedway, Austin, TX 78712, USA}

\author[0000-0002-0962-7585]{Mercier, S.}
\affiliation{Department of Earth, Atmospheric and Planetary Sciences,\\ Massachusetts Institute of Technology, Cambridge, MA 02139, USA}

\author[0000-0002-6892-6948]{Seager, S.}
\affiliation{Department of Physics and Kavli Institute for Astrophysics and Space Research, Massachusetts Institute of Technology, Cambridge, USA}
\affiliation{Department of Earth, Atmospheric and Planetary Sciences, Massachusetts Institute of Technology, Cambridge, USA}
\affiliation{Department of Aeronautics and Astronautics, Massachusetts Institute of Technology, Cambridge, USA}

\author[0000-0002-3418-8449]{Solanki, S.K.}
\affiliation{\mps}

\author[0000-0001-8217-6998]{Unruh, Y.}
\affiliation{Department of Physics, Imperial College London, UK}

\author[0000-0003-2415-2191]{de Wit, J.}
\affiliation{Department of Earth, Atmospheric and Planetary Sciences,\\ Massachusetts Institute of Technology, Cambridge, MA 02139, USA}

\author[0000-0002-0929-1612]{Witzke, V.}
\affiliation{University of Graz, Austria}


\begin{abstract}
Differences in the ingress and egress shapes of transit light curves can indicate morning-evening temperature contrasts on transiting planets. Here, we pinpoint an alternative mechanism that can introduce asymmetries in transit light curves, potentially affecting the accurate determination of morning-evening differences.
Small-scale magnetic field concentrations on the surfaces of the host star affect the visibility of stellar limb regions, making them brighter relative to the non-magnetic case. A difference in magnetization between the star's western and eastern limbs can thus create an asymmetry in limb brightness and, consequently, an asymmetry between transit ingress and egress. We model the limb darkening and stellar limb asymmetry in solar-like stars using the 3D radiative MHD code MURaM to simulate magnetized stellar atmospheres and the MPS-ATLAS code to synthesize spectra using ray-by-ray approach.
Our results show that ingress-egress depth differences can reach up to 600 ppm for a 10,000 ppm transit at 0.6 $\mu$m, depending on the magnetization of the stellar limbs—significantly interfering with planetary signals. Observations of the Sun show that such concentrations are often not accompanied by spots and do not manifest in photometric variability, indicating that even photometrically quiet stars can produce such asymmetries. However, planetary and stellar asymmetries exhibit distinct wavelength dependencies, which we propose to leverage for disentangling them.
\end{abstract}

%\keywords{stars: atmospheres --- stars: magnetic field --- (stars:) planetary systems --- Exoplanetary}


\section{Introduction} \label{sec:intro}
The James Webb Space Telescope \citep[JWST,][]{gardner_2006}, with its unprecedented precision in spectroscopic transit light curve measurements, offers the opportunity to study not only the chemical composition of exoplanetary atmospheres but also day-night differences within those atmospheres. The existence of such differences was predicted by \cite{Showman2002} soon after the discovery of the first exoplanet orbiting a solar-like star \citep{discovery}. They modeled atmospheric circulation on tidally locked planets and found morning-evening terminator temperature differences of up to 500~K. More recent studies have shown that cloud formation can amplify these differences \citep{2019A&A...631A..79H,2023A&A...671A.122H}.

The morning-evening terminator temperature contrasts could be observed through the differences between ingress and egress shapes of transit light curves. Despite the fact that \cite{Showman2002} proposed the idea more than two decades ago, the first observational sign of asymmetry using spectroscopic light curves was only reported very recently by \cite{espinoza2024} for transit observations of the WASP-39, which is approximately 400 K cooler than the Sun and has a near-solar metallicity \citep{Faedi2011}. They found about 400 ppm deeper transits in the near-infrared  from the evening part (egress) of the WASP-39b compared to the morning part (ingress) and estimated that this corresponds to about 200~K temperature contrast between evening and morning terminators. 

In this study we investigate whether the magnetic activity of a star can affect the measurements of the asymmetry between the morning and evening terminators. One of the most studied manifestations of stellar magnetic activity are bright and dark magnetic features on stellar surfaces \citep[e.g. spots and faculae,][]{Sami2006, Basri2021}.  Contamination by unocculted magnetic features have plagued transmission spectroscopy studies for the last decade and have been broadly discussed in the literature \citep[see, ][for a review]{2023Rackham}.

Here we point to another more intricate effect that introduces an asymmetry between ingress and egress shapes and can directly interfere with the measurements of morning-evening terminator temperature contrasts. It was recently demonstrated that the magnetic fields affect stellar limb darkening, making the stellar limb brighter relative to the non-magnetic case \citep{Ludwig2023, kostogryz2024}. Since the distribution of stellar surface magnetic field is rather inhomogeneous and asymmetric \citep{Reiners_2012} magnetic activity imposes the asymmetries in stellar limb darkening. The asymmetry is particularly pronounced at the limb, where magnetic effects are the strongest \citep{kostogryz2024}. (This phenomenon is readily apparent in solar white-light images, where concentrations of small-scale magnetic fields are visible only near the limb, see e.g. archive of \href{https://soho.nascom.nasa.gov/data/realtime/hmi_igr/512/}{SDO/HMI Continuum Images}). 

\begin{figure}
    \centering
\includegraphics[width=0.99\textwidth, trim=8cm 0cm 8cm 0cm, clip]{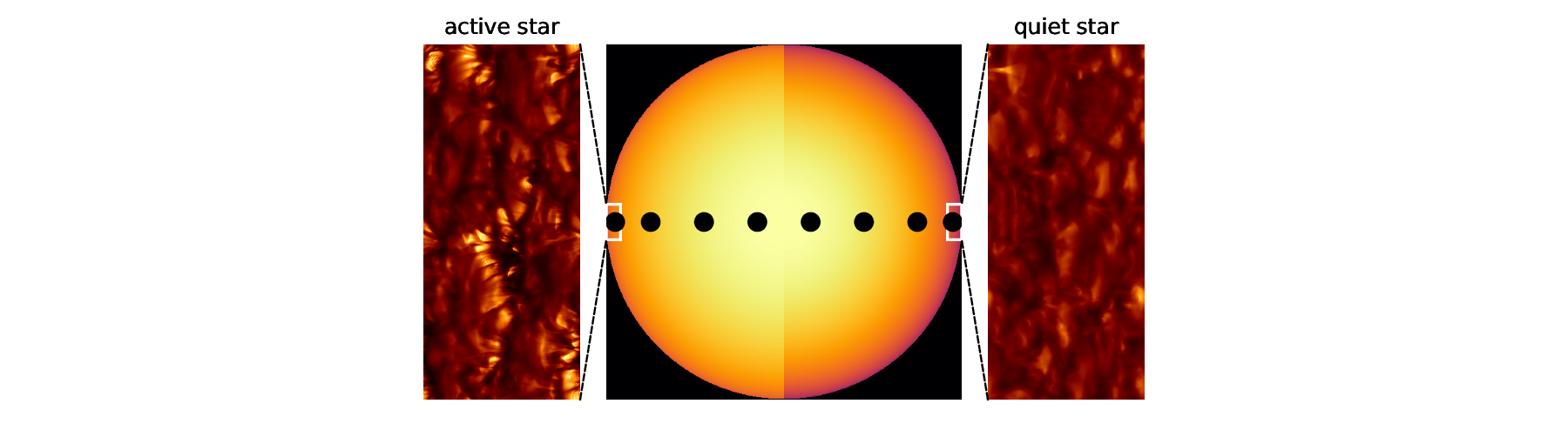}
    \caption{{\bf Sketch of a planet transiting a star with spatially inhomogeneous distribution of magnetic field.} The central panel shows a planet (black filled circles) transiting a star with active eastern hemisphere (left side of the star)  and quiet western hemisphere (right side). The effect of magnetic field on the stellar surface is especially pronounced at the near-limb stellar regions. We illustrate this by showing quiet (SSD, left panel) and active (300 G case, right panel)  MURaM snapshots that are computed with the MPS-ATLAS code at 600 nm for the near-limb viewing angle. The star, planet, and MURaM snapshots are not in scale. }
    \label{fig:muram_sim}
\end{figure}

The strong effect of magnetic fields on the brightness of the limb is caused by the corrugation of the magnetized stellar surface \citep{Solanki2013} and, thus, can best be reproduced by 3D MHD simulations. This is why the effect has been missed by the current generation of contamination models that are based on 1D static stellar atmospheric models. In contrast to these models, we base our calculations on simulations carried out by 3D radiative magnetohydrodynamics (MHD) code MURaM  ({\bf M}PS/{\bf U}niversity of Chicago {\bf Ra}diative {\bf M}HD) \citep{Voegler_2005, Rempel2014} that is capable of accurately  reproducing the magnetic effect on limb darkening \citep{kostogryz2024}.

\begin{figure}
    \centering
    \includegraphics[width=0.5\linewidth]{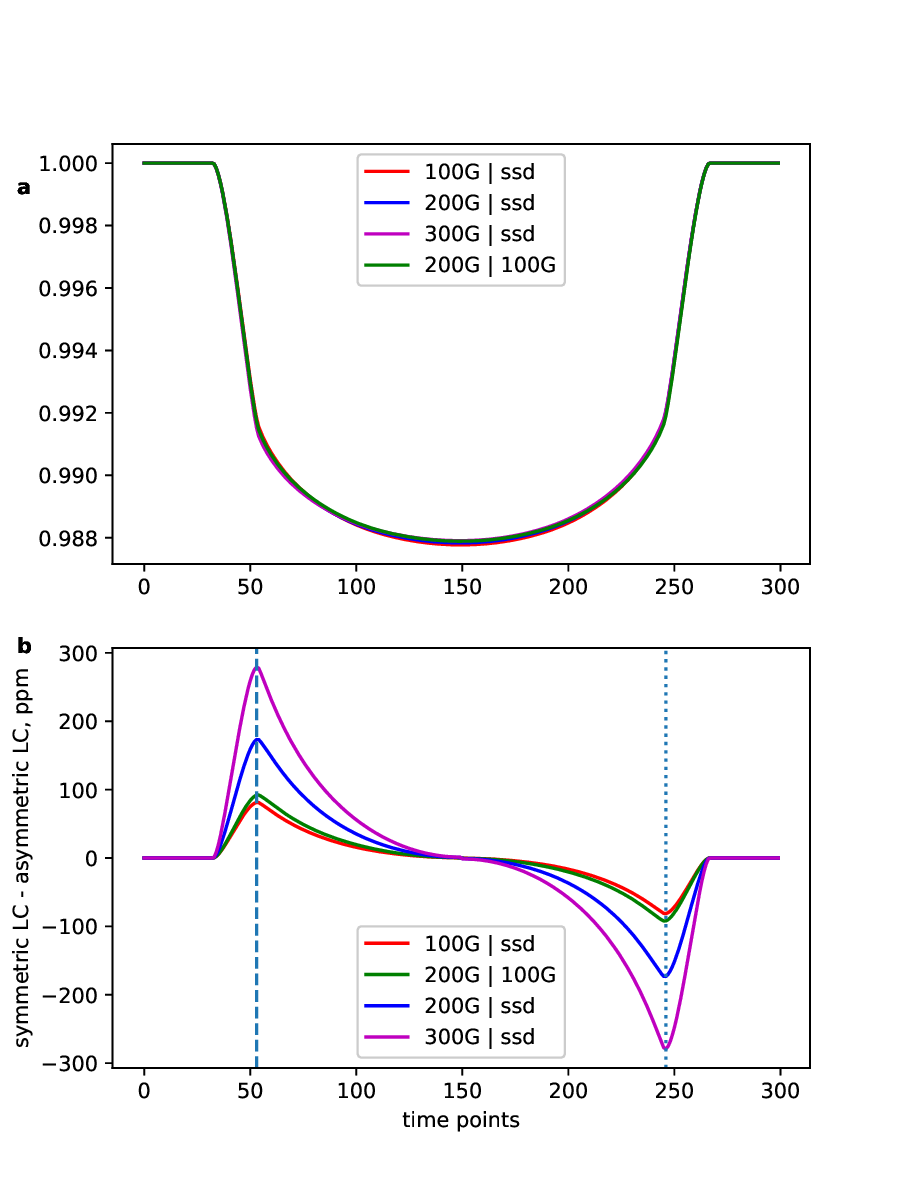}
    \caption{{\bf Asymmetry of the transit light curve caused by asymmetric spatial distribution of stellar magnetic activity.} Panel {\bf a} shows asymmetric light curves (flux normalised to unity outside of the transit as a function of time in arbitrary units) for a planet transiting a star, combined from half-disks with different magnetic activity. The different magnetic activity distributions are represented by lines of different colors: red, blue, and magenta correspond to a quiet eastern half-disk (represented by SSD) and 100 G, 200 G, and 300 G magnetization of the western half-disk, respectively. The green line corresponds to 200 G magnetization of the western half-disk and 100 G magnetization of the eastern half-disk. Panel {\bf b} illustrates the difference between the symmetric and asymmetric light curves. The light curve is simulated with a planet-to-star radius ratio of 0.1 and an impact parameter of 0.0 at a wavelength of 0.6 $\mu$m. The dashed vertical line in panel {\bf b} indicate the ingress transit depth, while the dotted line indicates the egress transit depth. The figure demonstrates that stellar surface magnetic fields have a pronounced effect on the shape of the transit ingress and egress. An asymmetric distribution of the magnetic field on the stellar disk causes asymmetry in the transit light curves. }
    \label{fig:limb_asymmetry}
\end{figure}

% \section{Model} \label{sec:model}
\section{Activity-induced asymmetry of transit light curves} \label{sec:model}
The main finding of this work is that stellar magnetic activity induces asymmetry between the ingress and egress of transit light curves. This effect is distinct from the widely acknowledged magnetic contamination of transmission spectra \citep{Rackham2017,Rackham2018,2023Rackham}. %stellar surface heterogeniety was already identified in M dwarfs \citep{Rackham2017,Rackham2018} as a significant contamination for regular exoplanet transit transmission observations, we find here that even G dwarfs can exhibit signification stellar contamination in ingress and egress transit observations.} 
The stellar asymmetry is driven by the small-scale concentrations of the magnetic field that manifest themselves as bright patches in the intergranular lanes. The ensembles of these patches form what is often referred to as photospheric faculae or bright spots. These patches are rare on quiet stellar surfaces (corresponding to minimal magnetic activity) (see Figure~\ref{fig:muram_sim}, right panel). Their presence becomes more pronounced in magnetically active cases (see Figure~\ref{fig:muram_sim}, left panel). The brightness and visibility of magnetic patches increase significantly towards the stellar limb, leading to a shallower limb darkening \citep{kostogryz2024}. 

The effect of magnetic field on stellar surface brightness have been obtained following the procedure and setup described by \cite{kostogryz2024}, who employed the 3D MHD MURaM code to simulate magnetic stellar atmospheres by 
solving the MHD equations using a 'box-in-a-star' approach. Namely, we simulated a small part of the stellar surface with the horizontal size of our simulations box is 9 Mm x 9 Mm and depth is 5 Mm \citep[see][]{muramisgood, kostogryz2024}. We represent the quiet stellar atmospheres by small-scale dynamo \citep[SSD;][]{Rempel2014, Rempel2018, bhatia_2022, witzke2022_SSD} simulations. The presence of ubiquitous field produced a small-scale turbulent dynamo is supported by solar observations \citep[e.g.,][]{Danilovic2010, Buehler2013}.
The magnetic activity produced by the global dynamo is simulated by introducing homogeneous vertical magnetic fields of 100~G, 200~G, and 300~G. These fields rapidly evolve into a statistically steady, inhomogeneous state with strong magnetic fields concentrated in intergranular lanes (forming bright patches) and much weaker fields in convective cells, which are mingled with those produced by the SSD that keeps running in the background. The spectral synthesis and the resulting limb darkening have been performed with the MPS-ATLAS code \citep{mps_atlas_2021} following the ray-by-ray approach \citep[see more details in][]{kostogryz2024}.

The magnetic field distribution on a stellar surface is not expected to be spatially homogeneous. Therefore, for any given transit, the magnetic field along the eastern and western near-limb portions (corresponding to ingress and egress, respectively) of the transit chord will differ, leading to a difference in brightness between them and consequently asymmetric transit light curves. We emphasize that the asymmetry arises from the inhomogeneous spatial distribution of the magnetic field on the stellar surface and is not related to its temporal evolution.  To illustrate this effect, we consider a simplified case of a star where eastern and western hemispheres have different magnetic activities and hence also levels of brightening introduced by its presence (see Figure~\ref{fig:muram_sim}). Each hemisphere is represented by a MURaM simulation with a given magnetic flux density (SSD, 100~G, 200~G, or 300~G). While a significant magnetization difference between entire eastern and western hemispheres (assumed here for simplicity) may be rare, the ingress and egress asymmetry is determined solely by the difference in magnetization between the near-limb regions. This difference can readily reach several hundred Gauss (see Sect.~\ref{sect:discussion}), so that our setup provides realistic estimation of ingress and egress asymmetry.

We simulate an equatorial transit with planet-to-star radius ratio ($r_p/R_\star$) of 0.1. The resulting light curves at $\lambda = 0.6 \mu m$ are shown in Figure~\ref{fig:limb_asymmetry}a. We refer to them as asymmetric light curves because their first and second halves are computed with different stellar magnetization levels. To pinpoint the asymmetry of these transit light curves, in Figure~\ref{fig:limb_asymmetry}b we show the difference between the asymmetric transit light curves and the symmetric light curves simulated with mean limb darkening averaged over both halves of the stellar disk. %These symmetric light curves represent are the best approximation of the light curves achievable with classic models assuming that planet and star are fully symmetric 
The  differences shown in Figure~\ref{fig:limb_asymmetry}b quantify the intrinsic limits of classic models assuming symmetric star and planet to describe the asymmetric transit light curves.
\begin{figure}
    \centering
    \includegraphics[width=0.8\linewidth]{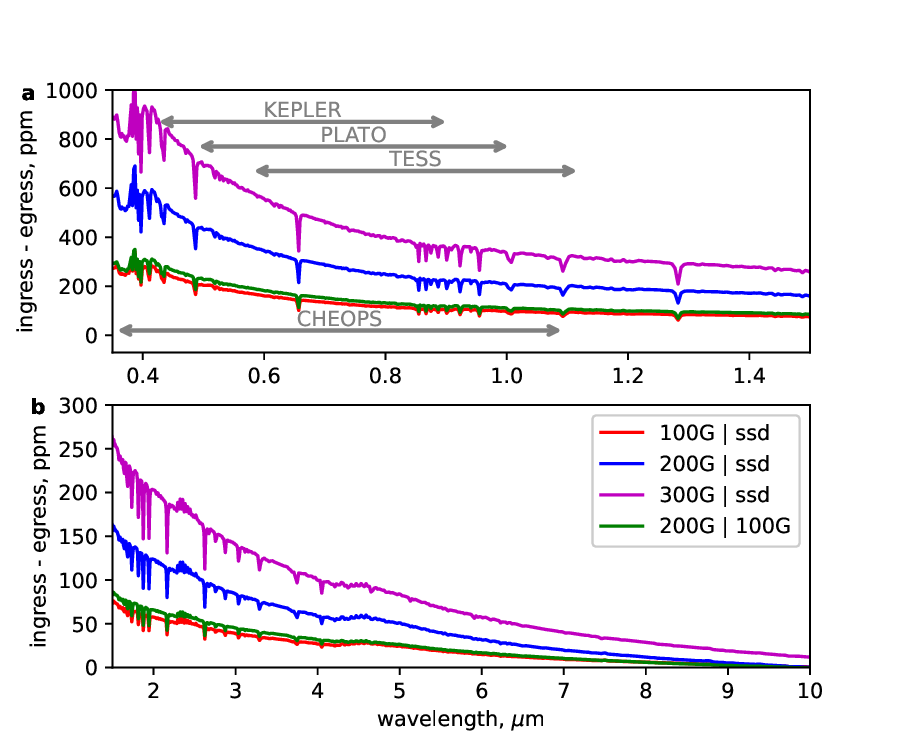}
    \caption{{\bf Magnetically-induced asymmetry of the transit light curve as a function of wavelength}. Plotted are the differences between transit depths of ingress and of egress at different wavelength ranges, i.e. from 0.35 to 1.5 $\mu m$ on panel {\bf a} and from 1.5 to 10 $\mu m$ on panel {\bf b}. The gray arrows demonstrate the passband width of different instruments: Kepler, PLATO, TESS, and CHEOPS.
    The light curve asymmetry is strongly wavelength-dependent, decreasing toward infrared wavelengths. % is presented for two wavelength ranges: from 0.4 to 2.0 $\mu$m in panels {\bf a} and {\bf b} for ingress and egress, respectively, and from 2.0 to 10.0 $\mu$m in panels {\bf c} and {\bf d} for ingress and egress, respectively. 
    }
    \label{fig:limb_asymmetry_spectrum}
\end{figure}

The maximum differences between mean and combined light curves occur right before the second and after third contacts (when the planet is nearly fully on the stellar disk). We remind  that the difference in transit depth between the second and third contacts is precisely the quantity used to assess the morning-evening temperature contrast on exoplanets \citep[][]{Showman2002, espinoza2024}. The asymmetry of the light curves increases (almost linearly) with the difference between magnetization levels of the eastern and western half-disks. At the same time it shows only marginal dependence on the absolute values of the magnetization (e.g., the red and green curves on Figure~\ref{fig:limb_asymmetry}b are almost identical). 

The asymmetry of the light curves shows a very pronounced dependence on the wavelength (see Figure~\ref{fig:limb_asymmetry_spectrum}). The amplitude of the asymmetry is highest in the visible wavelength range and then it decreases towards the infrared. This behavior is attributed to the decreasing sensitivity of the Planck function to a temperature change towards the infrared \citep{Seager2024}. The  'absorption' (Figure~\ref{fig:limb_asymmetry_spectrum}a) and 'emission' (Figure~\ref{fig:limb_asymmetry_spectrum}b) lines correspond to strong atomic lines (e.g. hydrogen lines are clearly visible) as well as to the molecular bands (e.g. CO bands at 2.2 $\mu$m and 4.5 $\mu$m).

\begin{figure}
    \centering
    \includegraphics[width=0.99\linewidth]{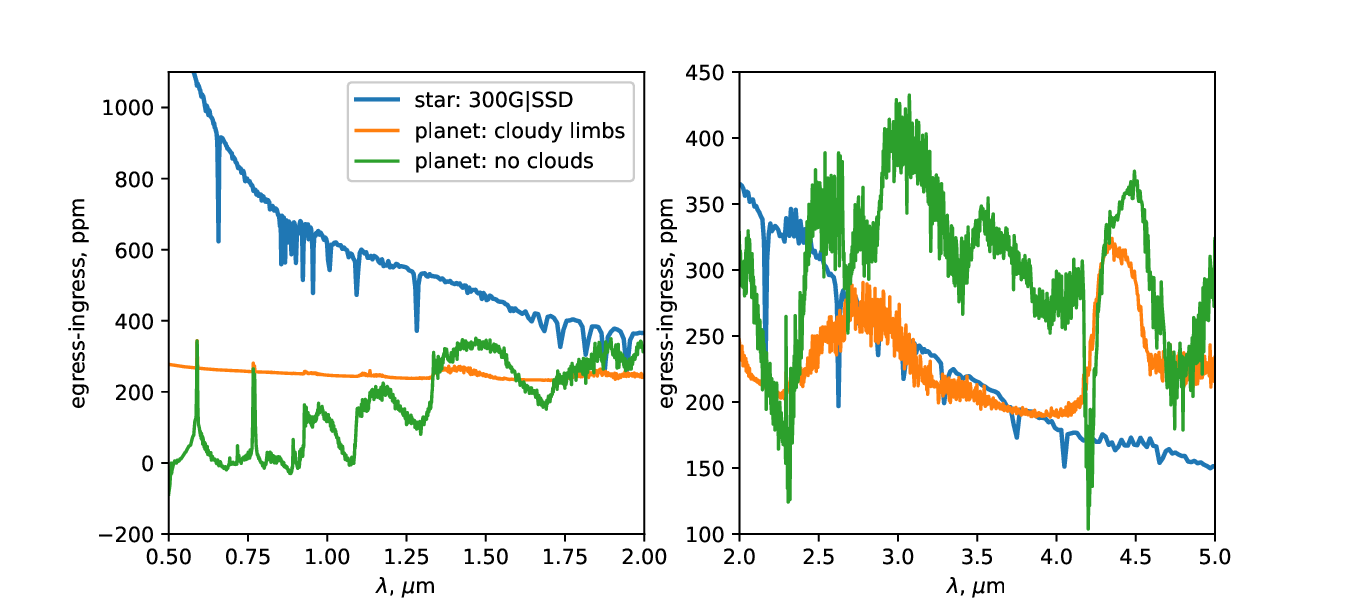}
    \caption{\textbf{Planet vs. stellar asymmetry for the illustrative case of  WASP-39}. Plotted are the differences between transit depths of ingress and egress as in Fig.~\ref{fig:limb_asymmetry_spectrum}. The blue line is the stellar magnetically-induced asymmetry for 300~G.  The orange and blue lines show planetary asymmetry following scenarios from the \texttt{IWF Graz 3D hierarchical cloud model}. 
    The wavelength dependencies of planetary and stellar signals differ significantly, providing a key to disentangle them.
    }
    \label{fig:planet_and_star_asymmetry}
\end{figure}

The wavelength dependencies of stellar and planetary asymmetries are distinctly different. We showcase it in the exemplary case of WASP-39b transits by overlaying stellar and planetary signals (Figure~\ref{fig:planet_and_star_asymmetry}). The stellar signal was computed using MURaM simulations of G9 ($T_{\rm eff}=5250$ K, $\rm M/H$=0.0, $\log g$ = 4.4) following the same setup as solar simulations \citep[][]{muramisgood}. We use impact parameter of 0.4 and planet-to-star radius ratio of 0.146.  We note, however, that the value of the impact parameter has no effect on ingress/egress asymmetry. 

The planetary signals are computed with the 3D climate model of \citet{Carone2023} in two setups: with and without clouds. These are the same planetary signals used by \citet{espinoza2024}, with the only exception being the omission of CH$_4$ as an opacity source in our study. This is because recent results indicate that CH$_4$ is vertically quenched in the WASP-39b atmosphere (Steinrueck et al., in prep).  We used the \texttt{Claus} extension of \texttt{petitRADTRANS} \citep{Kiefer2024,Molliere2019} to calculate cloudy evening and morning transit depths. 

The cloud-free case highlights asymmetry caused by the temperature difference between the evening and morning terminators, driven by circulation. Because the evening terminator is warmer than the morning terminator, the egress-ingress asymmetry is positive for all wavelengths.  The formation of clouds happens at both terminators (although evening terminator appears cloudier than the morning terminator) and significantly changes the planetary signal.
The clouds generally  mute spectral features in the asymmetry. It is apparent in the smaller egress-ingress signal of the cloudy model between 2--5 $\mu$m. The clouds also lead to a nearly wavelength-independent signal between 0.5--1.75 $\mu$m (although they generally increase the signal comparing to the cloud free case).

The main result from the comparison presented in Fig.~\ref{fig:planet_and_star_asymmetry} is that the stellar asymmetry can be comparable or even substantially larger than the planetary signal in the visible and near-IR. 
Nonetheless, there are ways of distinguishing between them based on their different wavelength dependencies, as the figure also shows. While the wavelength dependencies of stellar and planetary signals are similar between 2.5--4 $\mu$m, they are substantially different shortward 2.5 $\mu$m and at the 4.4~$\mu$m CO$_2$. Due to the different wavelength dependencies of stellar and planetary signals we do not expect that the signal discovered by \cite{espinoza2024} at 4.4~$\mu$m  can be solely explained by the magnetic activity of WASP-39. At the same time, our calculations show that magnetic activity can significantly contribute to transit asymmetries, even for relatively inactive G- and K-dwarfs. It can both amplify or mute such asymmetries (depending on which part of the stellar limb was more active during the transit). 

We emphasize that the stellar asymmetry shown in Fig.~\ref{fig:planet_and_star_asymmetry} is calculated for 300~G magnetization \citep[typical for spatially averaged fields in solar faculae,][]{Frazier1971} of the eastern limb and quiet western limb (see Fig.~\ref{fig:muram_sim}g). This can be considered as a somewhat extreme but still realistic scenario of a decaying active region at the eastern limb (see discussion in Sect.~\ref{sect:discussion}). The amplitude (but not the shape of the wavelength dependence) of the stellar signal will reduce if some activity is present on the western limb or if the eastern limb is less active (see Figure~\ref{fig:limb_asymmetry_spectrum}).

\section{Discussion and Conclusions}\label{sect:discussion}
We have shown that an asymmetric spatial distribution of magnetic activity (magnetic field) on a star 
leads to asymmetric transit light curves and can, thus, interfere with measuring morning-evening temperature contrasts on exoplanets. 
The contamination effect is caused by small-scale magnetic fields that increase the brightness of the stellar limb relative to the non-magnetic case. Unlike spots, which are compact and exhibit strong brightness contrast regardless of disk position, small-scale magnetic fields produce more diffuse, larger facular and network features. These features exhibit generally low brightness contrast except near the limb \citep[see, e.g., Figure 1 from][]{Witzke2022}, where the contrast strongly increases due to the effect of bright walls \citep{Solanki2013}. Therefore, they do not cause significant crossing events or stellar photometric variability (making them much harder to identify than spots), but, as we have demonstrated, can significantly contaminate measurements of morning-evening planetary asymmetry.

The presence of small-scale features on stellar surfaces does not necessarily implies the presence of spots. For example, on the Sun, spots typically live for days to a few weeks, while faculae decay on the time scale of several months \citep[see][for review]{Sami2006}. Consequently decaying active regions, which can reach the size of a typical gas giant planet (even on the relatively inactive Sun), are usually fully dominated by small-scale fields.

To distinguish stellar and planetary asymmetries with limited transit data, we recommend exploiting their distinct wavelength dependencies. With multiple transits, we anticipate the activity signal to diminish proportionally to the inverse square root of the number of observed transits. An important exception, however, is presented by non-equatorial transits with tilted orbits. In such cases, the transit chord can cross quiet stellar latitudes at one limb and active latitudes at the other, resulting in stellar asymmetry that cannot be averaged out over time. One additional source of the uncertainty in measuring transit asymmetry not considered in the present study is uncertainties in the determination of orbital parameters \citep[][]{deWit2012} due to the inaccurate treatment of the limb darkening \citep{Keers2024}. Following \cite{Keers2024} we recommend to use three-parameter law parameterizations of the limb darkening \citep{Sing2009, Kipping_3coeffs} for the determination of orbital parameters in all studies of transit asymmetry.

\begin{acknowledgments}

This work was supported by the ERC Synergy Grant REVEAL under the European Union’s Horizon 2020 research and innovation program (grant no. 101118581) and by German Aerospace Center (DLR) grants ``PLATO Data Center'' $50$OO$1501$ and $50$OP$1902$. We acknowledge support by the Max Planck Computing and Data Facility.
\end{acknowledgments}

% \appendix

% \section{Effect of impact parameter}

% We demonstrate here how limb asymmetry depends on the impact parameter (see Figure~\ref{fig:asymmetry_diff_b}). The difference between the ingress and egress is 600~ppm and remains constant across different impact parameters (see Figure~\ref{fig:asymmetry_diff_b} {\bf b} and {\bf d} for $b=0.4$ and $0.7$, respectively and Figure~\ref{fig:limb_asymmetry} {\bf b} for $b=0.0$). However the shape of the asymmetric profile varies between all three cases with $b=0, 0.4,$ and $0.7$. The jump in the middle of transit corresponds to a change in magnetization levels between the hemispheres. 

% \begin{figure}
%     \centering
%     \includegraphics[width=0.49\linewidth]{lc_and_diff_mean_b0.4_r0.1_wln600.eps}\
%     \includegraphics[width=0.49\linewidth]{lc_and_diff_mean_b0.7_r0.1_wln600.eps}
%     \caption{Asymmetric transit light curves similar to those plotted in Figure~\ref{fig:limb_asymmetry} (panels a and c, respectively) for impact parameters 0.4 and 0.7. Panels b and d: differences between asymmetric and symmetric transit light curves.}
%     \label{fig:asymmetry_diff_b}
% \end{figure}

\bibliography{references}{}
\bibliographystyle{aasjournal}

\end{document}